\documentclass{lmcs}

\usepackage{enumerate}
\usepackage{hyperref}
\usepackage{amssymb}

\theoremstyle{plain}
\newtheorem{example}[thm]{Example}

\newcommand\twoheaduparrow{\mathrel{\rotatebox{90}{$\twoheadrightarrow$}}}
\newcommand\twoheaddownarrow{\mathrel{\rotatebox{90}{$\twoheadleftarrow$}}}
\def\kurung#1{\left(#1\right)}
\def\himp#1{\{#1\}}

\def\real#1{\mathbb{#1}}

\def\mc#1{\mathcal{#1}}
\def\sub{\subseteq~\!\!\!}

\def\thdai{\twoheaddownarrow_{\Irr}}
\def\lli{\ll_{\Irr}}
\def\dua{\twoheaduparrow_{\Irr}}
\def\siconv{\xrightarrow{\Irr}}

\newcommand{\cl}{\operatorname{cl}}

\newcommand{\Irr}{\operatorname{Irr}}

\newcommand{\bigsup}{\bigvee}

\begin{document}

\title[On a new convergence class in $k$-bounded sober spaces]{On a new convergence class in $k$-bounded sober spaces}

\author[H. Andradi]{Hadrian Andradi}	
\address{National Institute of Education, Nanyang Technlogical University, 1 Nanyang Walk, Singapore 637616}	
\email{hadrian.andradi@gmail.com}  
\thanks{The first author is supported by Nanyang Technological University Research Scholarship (RSS)}	

\author[W. K. Ho]{Weng Kin Ho}	
\address{National Institute of Education, Nanyang Technlogical University, 1 Nanyang Walk, Singapore 637616}	
\email{wengkin.ho@nie.edu.sg}  




\begin{abstract}
 \noindent Recently, J. D. Lawson encouraged the domain theory community to consider the scientific program of developing domain theory in the wider context of $T_0$ spaces instead of restricting to posets.  In this paper, we respond to this calling by proving a topological parallel of a 2005 result due to B. Zhao and D. Zhao, i.e., an order-theoretic characterisation of those posets for which the lim-inf convergence is topological.  We do this by adopting a recent approach due to D. Zhao and W. K. Ho by replacing directed subsets with irreducible sets.  As a result, we formulate a new convergence class on $T_0$ spaces called $\Irr$-convergence and established that this convergence class $\mc{I}$ on a $k$-bounded sober space $X$ is topological if and only if $X$ is $\Irr$-continuous.
\end{abstract}

\maketitle

\section{Introduction}
\label{sec: intro}
Domain theory can be said to be a theory of approximation on partially ordered sets.
There are two sides of the same domain-theoretic coin: the order-theoretic one and the topological one.  On the order-theoretic side, the facility to approximate is built into the ordered structures via approximation relations, and here \emph{domains} is the generic term that includes all ordered structures that satisfy some approximation axioms.  On the topological side, approximation can be handled by \emph{topology}; more precisely, using net convergence.  Two famous results of D. S. Scott (\cite{scott72a}) epitomise this deep connection between domains and topology:  (1) A space is injective if and only if it is a continuous lattice with respect to its specialization order. (2) The lim-inf convergence class on a directed complete partial order (dcpo, for short) $P$ is topological if and only if $P$ is a continuous.  The second result was later generalised by B. Zhao and D. Zhao (\cite{zhaozhao05}) to the setting of posets which are not necessarily dcpo's.

In an invited presentation\footnote{This talk bears an extra-terrestrial title of ``\emph{Close Encounters of the Third Kind: Domain Theory Meets $T_0$-Spaces Meets Topology}".} at the 6th International Symposium in Domain Theory, J. D. Lawson gave further evidence from recent development in domain theory to illustrate this intimate relationship between domains and $T_0$ spaces.  In particular, it was pointed out that ``several results in domain theory can be lifted from the contexts of posets to $T_0$-spaces".  For example, (1) the topological technique of dcpo-completion of posets~\cite{zhaofan07} can be upgraded to yield D-completion of $T_0$ spaces (i.e., a certain completion of $T_0$ spaces to yield d-spaces) \cite{keimellawson09}, and (2) an important order-theoretic result known as Rudin's lemma~\cite{gierzetal83}, which is central to the theory of quasicontinuos domains, has a topological version \cite{heckmannkeimel13}.

In this paper, we respond (in a small way) to Lawson' call to develop the core of domain theory directly in topological spaces by establishing a topological parallel of the aforementioned result due to B. Zhao and D. Zhao (\cite[Theorem 2.1]{zhaozhao05}). In that paper, a lim-inf convergence defined on a poset using directed subsets is shown to be topological if and only if the poset is continuous.  To prove a topologically parallel result of this, we adopt the recent approach in~\cite{zhaoho15} by replacing directed subsets with irreducible subsets.  The motivation for their approach is based on the observation that the directed subsets of a poset are precisely its Alexandroff irreducible subsets.
Based on this replacement principle, we invent four topological analogues of the usual domain-theoretic notions:
(i)  a new way-below relation $\ll_{\Irr}$,
(ii) a new notion of $\Irr$-continuous space,
(iii) a new net convergence class $\mathcal{I}$ on a given topological space $X$, and
(iv) a new Scott-like topological space defined by irreducible sets.
The main result we obtained is:
\begin{thm} \label{thm: main thm}
The following are equivalent for a $k$-bounded sober space $X$:
\begin{enumerate}[\em(i)]
\item The net convergence class $\mathcal{I}$ on $X$ is topological.
\item $X$ is $\Irr$-continuous.
\end{enumerate}
\end{thm}
The notion of $k$-bounded sobriety which was first introduced in \cite{zhaoho15} as a generalisation of bounded sobriety (\cite{mislove99}) has close connections with the novel topology mentioned in (iv).  Because little is known about this kind of sobriety, it is one of the purposes of this paper to give a slightly better understanding of it in relation to net convergence.

We organise this paper in the following way. In Sections~\ref{sec: prelim} and \ref{sec: irr-cont spaces}, we summarise some of the recent results reported in~\cite{zhaoho15} that are essential in our ensuing development. These results concern the derived topology $SI(X)$ defined using irreducible sets of the underlying topology $X$, $k$-bounded sober spaces and $\Irr$-continuous spaces.  In Section~\ref{sec: convergence class defined by irreducible sets}, we introduce the new convergence class $\mc{I}$ defined on any given topological space $X$ and present some of its elementary properties.  Finally, we focus our development of the convergence class $\mc{I}$ on $k$-bounded sober spaces and prove the main characterisation theorem which we advertised in the abstract.

\section{Irreducibly derived topology}
\label{sec: prelim}
A nonempty subset $E$ of topological space $(X,\tau)$ is \emph{irreducible} if for any closed sets $A_1$ and $A_2$,
whenever $E \subseteq A_1 \cup A_2$, either $E \subseteq A_1$
or $E \subseteq A_2$.  The family of all irreducible subsets of $X$ is denoted by $\Irr_{\tau}(X)$ or $\Irr(X)$ whenever it is clear which topology one is referring to.

It is often useful to check the irreducibility of a set using open sets, i.e., $E$ is irreducible if and only if for any open sets $U_1$ and $U_2$, if $E \cap U_i \neq \emptyset$ ($i = 1,2$), then $E \cap U_1 \cap U_2 \neq \emptyset$.  Regarding irreducible sets, here are some elementary properties:
\begin{prop}
For any given topological space $X$, one has:
\begin{enumerate}[\em(1)]
\item $E \in \Irr(X)$ if and only if $\cl(E)\in \Irr(X)$.
\item The continuous image of an irreducible set is again irreducible.
\item If $\nu$ and $\tau$ are topologies on $X$ with $\nu\sub\tau$, then $\Irr_{\tau}(X)\sub\Irr_{\nu}(X)$.
\end{enumerate}
\end{prop}
A trivial example of irreducible set is a singleton.

Every $T_0$ space $(X,\tau)$ can be viewed as a partially ordered set via its \emph{specialisation order}, $\leq_\tau$, where $x \leq_{\tau} y$ if $x \in \cl_{\tau}(y)$.   For any subset $A$ of a $T_0$ space $X$, the supremum of $A$, denoted by $\bigsup A$,
always refers to the least upper bound of $A$ with respect to the specialisation order of $X$.  We denote the set of all irreducible subsets $X$ whose supremum exists by $\Irr^{+}(X)$.

A topological space $X$ is \emph{sober} if every irreducible closed set is the closure of a unique singleton.  All Hausdorff spaces are sober.  The Scott space of any continuous domain is sober.  A weaker form of sobriety is that of \emph{bounded sobriety} which requires that every irreducible closed set which is bounded above with respect to the specialisation order is the closure of a unique singleton. Bounded sober spaces have been studied in~\cite{mislove99} and~\cite{zhaofan07}. A yet weaker form of sobriety is that of $k$-bounded sobriety.  A topological space is \emph{$k$-bounded sober} if every closed set $F \in \Irr^+(X)$ is the closure of a unique singleton.  Every poset $P$ is $k$-bounded sober with respect to its upper topology, i.e., the coarsest one generated by sets of the form $P \backslash \downarrow x$,~$x \in P$.  With respect to its Scott topology, Johnstone's dcpo is not sober~\cite{johnstone81} but is $k$-bounded sober.  Indeed, all continuous posets are $k$-bounded sober with respect to the Scott topology.

Directed subsets play a central role in domain theory.  Directed subsets of a poset can be characterised topologically.  Recall that the Alexandroff topology on a poset $P$ consists of all upper sets.  The directed subsets of $P$ are precisely the Alexandroff irreducible subsets. The Scott topology is a coarsening of the Alexandroff topology in that every Scott open set is required to be upper and in addition inaccessible by directed suprema.  By replacing the directed sets by irreducible sets in the definition of a Scott open set, D. Zhao and W. K. Ho derived for any $T_0$ space (not just posets) a coarser topology called the \emph{irreducibly derived topology} that mimics the Scott topology on a poset.  More precisely, let $(X,\tau)$ be a $T_0$ space and $U \subseteq X$, define $U \in \tau_{SI}$ if
\begin{enumerate}[(1)]
\item $U \in \tau$, and
\item for every $E \in \Irr_{\tau}^{+}(X)$, $\bigsup E \in U$ implies $E \cap U \neq \emptyset$.
\end{enumerate}
It can be easily verified that $SI(X) := \kurung{X,\tau_{SI}}$ is a topological space coarser than $(X,\tau$).

Because the Scott-like topology $\tau_{SI}$ is derived from an underlying topology $\tau$ on the same set $X$, we sometimes refer to $\tau_{SI}$ as the \emph{Scott derivative} of $\tau$.

\begin{prop} \cite{zhaoho15}
Let $(X,\tau)$ be a $T_0$ space. Then the following hold:
\begin{enumerate}[\em(1)]
\item For any $x \in X$, $\cl_X(\himp{x}) = \cl_{SI(X)}(\himp{x})$.
\item A subset $C$ of $X$ is closed in $SI(X)$ if and only if for every $E \in\Irr_{\tau}^{+}(X)$, $E \subseteq C$ implies $\bigsup E \in C$.
\item An open subset $U$ of $(X,\tau)$ is $SI(X)$-open if and only if for any $E \in \Irr_{\tau}^{+}(X) \cap \Gamma(X)$, $\bigsup E \in U$ implies $E \cap U \neq\emptyset$.
\item A subset $U$ of $X$ is clopen in $X$ if and only if it is clopen in $SI(X)$.
\item $X$ is connected if and only if $SI(X)$ is connected.
\end{enumerate}
\end{prop}

\begin{example}\label{ex:si_topology}\hfill
\begin{enumerate}[\em(1)]
\item For any indiscrete space $X$, the $SI(X)$ is itself.
\item Let $P$ be a poset endowed with Alexandroff topology $\Upsilon(P)$. Since the irreducible sets are precisely the directed ones, it is clear that $SI(\Upsilon(P)) = \Sigma(P)$, where $\Sigma(P)$ is the Scott topology on $P$.
\end{enumerate}
\end{example}

In general, the Scott topology of a given poset does not coincide with its Alexandroff topology.   For example, in the set $\real{R}$ of all real numbers equipped with the usual order,  sets of the form $[x,\infty)$ are Alexandroff open but not Scott open. In general, any $T_0$ space is strictly coarser than its Scott derivative.  We shall now look at those spaces which are equal to their Scott derivatives.

Let $(X,\tau)$ be a $T_0$ space and $\alpha$ an ordinal.
We define by transfinite induction a topological space $X^{\alpha}$ on $X$ as follows:
\begin{enumerate}[(1)]
\item $X^{0} := (X,\tau)$;
\item $X^{\alpha + 1} := SI\kurung{X^{\alpha}}$;
\item If $\alpha$ is a limit ordinal, then $X^{\alpha}$ is the space on $X$ whose topology is the intersection of all topologies $X^\beta$, where $\beta < \alpha$.
\end{enumerate}
Since $(X_\alpha)_\alpha$ is a sequence of increasingly coarser
topologies on $X$, there is a smallest ordinal $\gamma(X)$ such that topology on $X^{\alpha}$ coincides with that on $X^{\gamma}$ for all $\alpha\geq \gamma(X)$. We denote this $X^{\gamma(X)}$ by $X^{\infty}$.

A topological space $(X,\tau)$ is said to satisfy the $SI^{\infty}$ \emph{property} if $\tau = \tau_{SI}$.  A natural question is to ask for a characterisation of spaces which satisfy the
$SI^\infty$ property.  It turns out that for any topological space $(X,\tau)$ we have:
\begin{thm} \cite[Theorem 4.5]{zhaoho15}
\label{thm: char of kb-space}
$X$ is $k$-bounded sober if and only if $X$ satisfies the $SI^{\infty}$ property.
\end{thm}

\section{$\Irr$-continuous spaces}
\label{sec: irr-cont spaces}
In a topological space $(X,\tau)$, one defines a ``new" way-below relation $\lli$ (called the \emph{$\Irr$-way-below relation}) using the irreducible subset instead of directed subset.
Given $x, y \in X$, we define
\[
x \lli y \iff \forall E \in \Irr^{+}(X).~(\bigsup E \geq_\tau y) \implies (E \cap \uparrow x \neq \emptyset).
\]
For a given $x \in X$, $\thdai x$ denotes the set $\himp{y \in X \mid y \lli x}$.
The following properties of $\Irr$-way-below relation are to be expected:
\begin{prop}
\label{prop: transitivity of lli}
In a space $X$ the following hold for all $u,x,y$ and $z \in X$:
\begin{enumerate}[\em(1)]
  \item $x \lli y$ implies $x \leq y$.
  \item $u \leq x \lli y\leq z$ implies $u \lli z$.
\end{enumerate}
\end{prop}

Using $\lli$, we can now introduce the notion of $\Irr$-continuous space -- a topological analogue of continuous posets.
\begin{defi}
A topological space $(X,\tau)$ is said to be $\Irr$\emph{-continuous} if for every $x\in X$ the following hold:
\begin{enumerate}[(1)]
\item $\thdai x$ is irreducible in $(X,\tau)$ and
\item $x = \bigsup \thdai x$.
\end{enumerate}
\end{defi}

\begin{rem}
Our definition of $\Irr$-continuous space differs from that of SI-continuous spaces defined in \cite[p.192]{zhaoho15} in that we choose to drop a third condition:
\begin{enumerate}[(3)]
\item  For any $x \in X$, $\dua x$ is open in $(X,\tau)$.
\end{enumerate}
Our choice of omission is deliberate because of a result by M. Ern\'{e} (\cite[Theorem 4, p.462]{erne05}).  That result asserts that a topological space is a weak C-space (i.e., it is both a C-space and a weak monotone convergence space) if and only if it is homeomorphic to the Scott space of some continuous poset.  It was shown in~\cite[Theorem 6.4]{zhaoho15} that $X$ is SI-continuous if and only if the derived topology $SI(X)$ is a C-space.  Because $SI(X)$ it always a weak monotone convergence space, it follows that the derived topology on an SI-continuous space is homeomorphic to the Scott topology on some continuous poset.
However, this fact will go contrary to our original intention of developing domain theory in a wider contexts of topological spaces and \emph{not restricted just to (continuous) posets}.  Thus, we must take out this third condition from the definition of SI-continuity to formulate our present definition of $\Irr$-continuity.
\end{rem}

With the absence of the third condition, we can still say a few things about $\Irr$-continuous spaces in general.
\begin{lem}
\label{lem: prelude to interpolating}
Let $X$ be an $\Irr$-continuous space.  Then, for every $x \in X$ it holds that
\[
x=\bigsup \bigcup \himp{\thdai y\mid y\lli x}.
\]
\end{lem}
\proof
Let $M_x := \bigcup \himp{\thdai y\mid y\lli x}$. It is clear that $x$ is an upper bound of $M_x$. We shall show that $u \geq x$ for any upper bound $u$ of $M_x$. Suppose for the sake of contradiction that $u \ngeq x$.  Then, by the $\Irr$-continuity of $X$, $x = \bigsup \thdai x$ so that there exists $y \in \thdai x$ with $y \nleq u$.  Repeating the same argument we can find a $z \in \thdai y$ such that $z \nleq u$.  But this is a contradiction to the fact that $z \in M_x$ and $u$ is an upper bound of $M_x$.  Therefore, $u \geq x$ and this completes the proof.\qed

Any domain theorist would know the price for dropping the third condition, i.e., one loses the \emph{interpolating property} of the way-below relation.  Fortunately, within the scope of our present study concerning $k$-bounded sober spaces, we can recover this loss.

\begin{thm}\label{thm: interpoting prop in SI infty}
Let $X$ be an $\Irr$-continuous and $k$-bounded sober space.
Then, $\lli$ enjoys the interpolating property in that whenever $z\lli x$, there exists $y \in X$ such that
\[
z \lli y\lli x.
\]
\end{thm}
\proof
We first show that $M_x := \bigcup \himp{\thdai y\mid y\lli x}$ is an irreducible subset in $(X,\tau)$. Let $U_1$ and $U_2$ be open sets in $X$ such that $M_x\cap U_1\neq\emptyset$ and $M_x\cap U_2\neq\emptyset$. Then there exist $y_1,y_2\in \thdai x$ such that $y_1\in U_1$ and $y_2\in U_2$. Since $x$ is an upper bound of $\himp{y_1,y_2}$ and both $U_1$ and $U_2$ are upper sets, $x\in U_1\cap U_2$.  By $\Irr$-continuity of $X$, $x$ is the supremum of $\thdai x$. Since $X$ is $k$-bounded sober, it enjoys the $SI^{\infty}$ property and so $U_1,~U_2\in SI(X)$.  Hence there exists $y \in \thdai x$ such that $y \in U_1 \cap U_2$. Using a similar argument, there exists $z \in \thdai y$ such that $z \in U_1 \cap U_2$. Therefore, there exists $z \in X$ such that $z \in M_x \cap U_1 \cap U_2$. Consequently, $M_x$ is an irreducible subset of $X$.  Now, let $z \lli x$.  Since $M_x$ is $\tau$-irreducible and, by Lemma~\ref{lem: prelude to interpolating}, $\bigsup M_x = x$, there exists $w \in M_x$ such that $z \leq w$. Hence there exists $y \in X$ such that, by virtue of Proposition~\ref{prop: transitivity of lli}, $z \lli y \lli z$ holds as desired. \qed

\begin{example}
The rational line $Q := (\mathbb{Q},\leq)$ with the Scott topology $\Sigma Q$ is an $\Irr_\Sigma$-continuous and $k$-bounded sober space.
\end{example}

\section{Convergence class defined by irreducible sets}
\label{sec: convergence class defined by irreducible sets}
In a topological space, approximation can be described by means of net convergence.
Let $X$ be a set.  A \emph{net} $(x_i)_{i \in I}$ in $X$ is a mapping from a preorder $(I,\leq)$ to $X$.  Real number sequences, for instance, are nets in the Euclidean space $\mathbb{R}$.  Thus, nets can be viewed as generalised sequences.  We denote the set of all nets in $X$ by $\Psi X$.

For each $x \in X$, one can define a \emph{constant net} by $x_i = x$ for all $i \in I$.  Parallel to the notion of subsequence, we have the notion of a subnet.
A net $\left(y_k\right)_{k\in K}$ is a \emph{subnet} of $\left(x_j\right)_{j\in J}$ if (i) there exists a monotonic function $f:K \to J$ such that $y_{k} = x_{f(k)}$ for all $k \in K$ and (ii) for each $j \in J$ there exists $k_j \in K$ with $f(k_j)\geq j$.

A convergence class $\mathcal{S}$ on a set $X$ is a relation between $\Psi X$ and $X$.
An element of $\mathcal{S}$ is denoted by $((x_i)_{i \in I},x)$ or sometimes
$(x_i)_{i \in I} \xrightarrow{\mathcal{S}} x$, in which case we say that the net
$(x_i)_{i \in I}$ \emph{$\mathcal{S}$-converges} to $x$.

Every topological space $(X,\tau)$ induces a convergence class
$\mathcal{S}_\tau$ defined by
\[
(x_i)_{i \in I} \xrightarrow{\mathcal{S}_\tau} x \iff
\forall U \in \tau.~(x \in U) \implies x_i \in U \text{ eventually.}
\]
Here, a property of a net $(x_i)_{i \in I}$ holds eventually if there exist $i_0 \in I$ such that for all $i \geq i_0$, the property holds for $x_i$.

When $(x_i)_{i \in I} \xrightarrow{\mathcal{S}_\tau} x$, we say that $(x_i)_{i \in I}$ converges topologically to $x$.  A convergence class, $\mathcal{S}$, on a set $X$, is said to be \emph{topological} if there is a topology $\tau$ on $X$ that induces it, i.e., $\mathcal{S} = \mathcal{S}_{\tau}$.

A special convergence class on a dcpo called the \emph{lim-inf convergence} was first introduced in~\cite{scott72a}. Crucially, this convergence makes use of the directed sets.
It was shown that the lim-inf convergence on a dcpo is topological if and only if the dcpo is a domain.  Later in~\cite{zhaozhao05}, this lim-inf convergence was modified to create a new convergence class for a general poset.  Recall that in a poset $P$, a net $\kurung{x_i}_{i\in I}$ converges to $y$ provided that there exists a directed subset $D$ of eventually lower bounds of $\kurung{x_i}_{i\in I}$ whose supremum belongs to $\uparrow y$. In that later paper, it was established that the new lim-inf convergence on a poset is topological if and only if the poset is continuous.

In this paper, we modify the preceding definition of convergence to suit the context of a topological space by replacing the directed subsets with irreducible subsets.
\begin{defi}
\label{def: Irr-convergence}
Let $(X,\tau)$ be a topological space.  A net $\kurung{x_i}_{i\in I}$ in $(X,\tau)$
is said to \emph{$\Irr$-converge} to $y \in P$ if there exists $E \in \Irr^+(X)$ such that $\bigsup_\tau E \geq_\tau y$ and for each $e \in E$ there exists $k(e) \in I$ such that for all $i \geq k(e)$ it holds that $x_i \geq_\tau e$. An instance of $\kurung{x_i}_{i \in I}$ converging to $x$ is denoted by $\kurung{x_i}_{i \in I} \siconv y$.
\end{defi}
Equivalently, $(x_i)_{i \in I} \siconv y$ if and only if there exists an irreducible subset $E$ of eventually lower bounds of $\kurung{x_i}_{i\in I}$ and whose supremum exists and belongs to $\uparrow y$. When the context is clear, the specialisation order (and the supremum taken with respect to it) shall be written as $\leq$ (and $\bigsup$), suppressing the subscripts.

The convergence class on a topological space $X$ defined by $\siconv$ is denoted by $\mathcal{I}$.  The rest of this section is completely devoted to studying $\mathcal{I}$; in particular, we obtained a necessary and sufficient condition on $X$ for which the convergence class $\mathcal{I}$, where $X$ is any $k$-bounded sober space, is topological.

The following two results characterise $\lli$ in terms of the convergence $\siconv$.
\begin{prop} \label{prop: lli char}
Let $X$ be a topological space and $x, y \in X$.
Then $x \lli y$ if and only if for any net $\kurung{x_i}_{i\in I} \siconv y$, there exists $k \in I$ such that $x_i \geq x$ for all $i\geq k$.
\end{prop}
\proof
Assume that $x \lli y$ and suppose that $\kurung{x_i}_{i\in I} \siconv y$. Then one can find an irreducible set $E$ such that $y \leq \bigwedge E$ and for each $e \in E$ there exists $k(e) \in I$ such that $x_i \geq e$ for all $i \geq k(e)$. Using the fact that $x \lli y$, there exists $e \in E$ such that $x \leq e$. Hence for all $i \geq k(e)=:k$ it holds that $x_i \geq e$.

For the converse, let $E$ be an irreducible set with $\bigsup E \geq y$. We must show that there exists $e \in E$ such that $e \geq x$.  Write $E = \{x_i \mid i \in I\}$ and preorder
the index set $I$ as follows: $i \leq_I j$ if and only if $x_i \leq x_j$.  This makes $(x_i)_{i \in I}$ a net in $X$.
We now show that $(x_i)_{i \in I} \siconv y$.  To this end, we claim that the irreducible set $E$ will satisfy the condition in Definition~\ref{def: Irr-convergence}.  Of course, $\bigsup E \geq y$ holds by assumption.  Now, for any $e \in E$, there exists $i \in I$ such that $x_i = e$.  If $j \geq i$, then by definition of the preorder on $I$, we have $x_j \geq x_i = e$.
Thus, $x \lli y$ as desired.\qed

\begin{lem}\label{lem: lli char}
Let $\kurung{x_i}_{i\in I}$ be a net in an $\Irr$-continuous space $X$, and $y \in X$.
Then $x_i\siconv y$ if and only if for each $x \in \thdai y$, there is $k(x) \in I$ such that for each $i \geq k(x)$ it holds that $x_i \geq x$.
\end{lem}
\proof
Let $(x_i)_{i \in I} \siconv y$ and $x \lli y$. By Proposition~\ref{prop: lli char}, there exists $k(x) \in I $ such that $x_i \geq x$ for every $i \geq k(x)$.

Conversely, we have that $\thdai y$ is irreducible in $X$ and $\bigsup \thdai y = y$. The assumption asserts that for each $x\in \thdai y$ there is $k(x) \in I $ such that if $i \geq k(x)$ then $x_i \geq x$. Therefore, $(x_i)_{i \in I} \siconv y$.\qed

From \cite{kelley55}, we know that a convergence class $\mc{S}$ is topological if and only if it satisfies the following conditions:
\begin{enumerate}[(1)]
\item (Constants). If $\kurung{x_i}_{i\in I}$ is a constant net with $x_i=x$ for all $i$, then $\kurung{\kurung{x_i}_{i\in I},x}\in \mc{S}$.
\item (Subnets). If $\kurung{\kurung{x_i}_{i\in I},x}\in\mc{S}$ and $\kurung{y_j}_{j\in J}$ is a subnet of $\kurung{x_i}_{i\in I}$, then $\kurung{\kurung{y_j}_{j\in J},x}\in \mc{S}$.
\item (Divergence). If $\kurung{\kurung{x_i}_{i\in I},x}\notin \mc{S}$, then there exists a subnet $\kurung{y_j}_{j\in J}$ of $\kurung{x_i}_{i\in I}$ such that for any subnet $\kurung{z_k}_{k\in K}$ of $\kurung{y_j}_{j\in J}$, $\kurung{\kurung{z_k}_{k\in K},x} \notin \mc{S}$.
\item (Iterated limits). If $\kurung{\kurung{x_i}_{i\in I},x}\in \mc{S}$ and $\kurung{\kurung{x_{i,j}}_{j\in J(i)},x_i} \in \mc{S}$ for all $i\in I$, then $\kurung{\kurung{x_{i,f(i)}}_{(i,f)\in I\times M},x} \in \mc{S}$, where $M :=\prod\himp{J(i)\mid i\in I}$.
\end{enumerate}
We shall rely on this result in characterising those $k$-bounded sober spaces $X$ for which $\mathcal{I}$ is topological.

\begin{lem}\label{lem:SI-class_is_topological}
Let $X$ be a $T_0$ space. 
\begin{enumerate}[\em(1)]
\item The class $\mc{I}$ satisfies the axioms (Constants) and (Subnets).
\item If $X$ is $\Irr$-continuous, then $\mc{I}$ satisfies the (Divergence) axiom.
\item If $X$ is $\Irr$-continuous and $k$-bounded sober, then $\mc{I}$ satisfies the (Iterated limits) axiom.
\end{enumerate}
\end{lem}
\proof\hfill
\begin{enumerate}[(1)]
\item That $\mc{I}$ satisfies the (Constants) axiom is immediate. We now show that
$\mc{I}$ satisfies the (Subnets) axiom.  Let $\kurung{\kurung{x_i}_{i\in I},x}\in \mc{I}$. Then there exists an irreducible subset $E$ of $X$ such that $x \leq \bigsup E$ and for each $e \in E$ there exists $k(e)\in I$ satisfying $x_i \geq e$ for all $i \geq k(e)$.
Let $\kurung{y_j}_{j\in J}$ be a subnet of $\kurung{x_i}_{i\in I}$, with $y_j = x_{g(j)}$ for each $j$. Then there exists $j'(e)\in J$ such that $g(j'(e)) \geq k(e)$. For all $j \in J$ such that $j \geq j'(e)$, it holds that $g(j)\geq g(j'(e))$, hence $y_j = x_{g(j)}\geq e$. Therefore, $\kurung{\kurung{y_j}_{j\in J},x} \in \mc{I}$.

\item Suppose $\kurung{\kurung{x_i}_{i\in I},x} \not \in \mc{I}$. By virtue of the $\Irr$-continuity of $X$, $\thdai x$ is an irreducible subset of $X$ and $\bigsup \thdai x = x$.  So there exists $y \in \thdai x$ such that for each $i \in I$ one can find $j(i) \in I$ satisfying $j(i) \geq i$ and $x_{j(i)} \ngeq y$.
    Define $J := \himp{j \in I \mid x_j \ngeq y}$.
    Then $\kurung{x_j}_{j \in J}$ is a subnet of $\kurung{x_i}_{i \in I}$.
    For every subnet $\kurung{z_k}_{k\in K}$ of $\kurung{x_j}_{j\in J}$ we have that $z_k \ngeq y$.
    By Lemma~\ref{lem: lli char}, $\kurung{\kurung{z_k}_{k\in K},x}$ cannot belong to $\mc{I}$. Thus, $\mc{I}$ satisfies the (Divergence) axiom.

\item We now prove that $\mc{I}$ satisfies the (Iterated limits) axiom.
Let $\kurung{\kurung{x_i}_{i\in I},x} \in \mc{I}$ and
$\kurung{\kurung{x_{i,j}}_{j\in J(i)},x_i} \in\mc{I}$ for all $i \in I$.
Let $y \lli x$. Since $X$ is $\Irr$-continuous and $k$-bounded sober, by Corollary~\ref{thm: interpoting prop in SI infty} there exists $z \in X$ such that $y \lli z\lli x$. Applying Lemma~\ref{lem: lli char} to the situation where $x_i \siconv x$ and $z \lli x$, there exists $k(z)\in I$ such that $x_i \geq z$ for all $i \geq k(z)$.  We then have $y \lli x_i$ for all such $i$. Similarly, applying Lemma~\ref{lem: lli char} to the situation where $x_{i,j} \siconv x_i$ and $y \lli x_i$, there exists $g(i) \in J(i)$ such that if $j \geq g(i)$ then $x_{i,j} \geq y$.

Let $M := \prod\himp{J(i)\mid i\in I}$.
Define $h \in M$ such that $h(i) = g(i)$ if $i \geq k(z)$ and $h(i)$ is any element in $J(i)$, otherwise. If $(i,f) \in I \times M$ with $(i,f) \geq (k(z),h)$, then $f(i) \geq h(i) \geq g(i)$, hence $x_{i,f(i)} \geq z$. By Theorem~\ref{thm: interpoting prop in SI infty}, $x_{i,f} \siconv x$. Therefore, $\mc{I}$ satisfies the (Iterated limits) axiom. \qed
\end{enumerate}

\begin{lem}\label{lem: subset thadai x}
Let $X$ be a $k$-bounded sober space and $x \in X$. If $E$ is an irreducible subset of $X$ such that $\bigsup E \geq x$ and $E \subseteq \thdai x$, then $\thdai x$ itself is irreducible in $X$.
\end{lem}
\proof
Let $U_1$ and $U_2$ be open in $X$ such that $\thdai x~ \cap U_1 \neq \emptyset$ and $\thdai x~ \cap U_2 \neq \emptyset$. Then there exist $w_k \in X$ ($k =1,2$) such that $w_k \lli x$ and $w_k \in U_k$.  Since $U_1$ and $U_2$ are upper, we have $x \in U_1 \cap U_2$. Further, since $\bigsup E \geq x$ we have $\bigsup E \in U_1 \cap U_2$. Since $X$ is $k$-bounded sober, $U_1 \cap U_2$ is open with respect to $SI(X)$.  So, there exists $e \in E$ such that $e \in U_1 \cap U_2$. By assumption $E \subseteq \thdai x$ so that $e \in E$ implies $e \lli x$.
Hence $\thdai x \cap U_1 \cap U_2$ is nonempty. So, $\thdai x$ is irreducible in $X$.\qed

\begin{lem}\label{lem:iter_lim_implies_SIcts}
For any $k$-bounded sober space $X$, if $\mc{I}$ satisfies the (Iterated limits) axiom then $X$ is $\Irr$-continuous.
\end{lem}
\proof
Let $x \in X$ and $\mc{F}_x=\himp{\kurung{x_{i,j}}_{j\in J(i)}\mid i\in I}$ be the family of all irreducible subsets of $X$ whose supremum exists and is greater than or equal to $x$.

For each $i\in I$, let $x_i := \sup \himp{x_{i,j}\mid j\in J(i)}$.
Then $x_i \geq x$ for all $i$. Since the constant net $(x) \in \mc{F}_x$, we have $\inf \himp{x_i \mid i\in I} = x$. We equip the index set $I$ with preorder $\leq$ such that $i_1 \leq i_2$ for any two $i_1,i_2\in I$.  We then have $(x_i)_{i \in I} \siconv x$; just take $\himp{x}$ as the irreducible set satisfying the definition.

For all $i \in I$, define a preorder $\leq$ on $J(i)$ as follows: $j_1 \leq j_2$ if and only if $x_{i,j_1} \leq x_{i,j_2}$.  We then have $(x_{i,j}) \siconv x_i$; just take $\himp{x_{i,j}\mid j\in J(i)}$ as the required irreducible set.

Let $M := \prod \himp{J(i) \mid i\in I}$. By assumption, we have that the net $\kurung{x_{i,f(i)}}_{(i,f) \in I \times M} \siconv x$. Thus, we can find an irreducible set $E$ such that
\begin{enumerate}[(1)]
\item $\bigsup E \geq x$ and
\item for each $e \in E$, $x_{i,f(i)} \geq e$ eventually.
\end{enumerate}
We now show that $E \subseteq \thdai x$.
Let $e \in E$.  For any irreducible set $K$ with $\bigsup K \geq x$,
$K = \himp{x_{i_0,j}\mid j\in J(i_0)}$ for some $i_0 \in I$.
For this $e \in E$, because $x_{i,f(i)} \geq e$ eventually, there exists $(i_e,f_e)$
such that if $(i,f) \geq (i_e,f_e)$, then $x_{i,f(i)} \geq e$.
By the definition of the preorder defined on $I$, $i_0 \geq i_e$ holds.
Hence $x_{i_0,f(i_0)} \geq e$.  Since $x_{i_0,f(i_0)} \in K$, it follows that
$e \lli x$. Thus, $E$ is an irreducible subset such that $E \subseteq \thdai x$ and $\bigsup E = x$, and so by Lemma~\ref{lem: subset thadai x}, $\thdai x$ is irreducible.
By virtue that $X$ is $k$-bounded sober, $\cl(\thdai x) = \cl(\{x\})$, whence $\bigsup \thdai x = x$.  So, $X$ is $\Irr$-continuous.
\qed

Finally, our main result, i.e., Theorem~\ref{thm: main thm}, is an immediate consequence of Lemmas~\ref{lem:SI-class_is_topological} and \ref{lem:iter_lim_implies_SIcts}.

\section{Conclusion}
\label{sec: conclusion}
In this paper, we take a small step towards taking up the programme of exporting domain theory to the more general context of a $T_0$ space.  The key strategy involved in our approach is to simply \emph{replace} directed subsets by irreducible sets -- a methodology first introduced by Zhao and Ho~\cite{zhaoho15}.  Recently, the importance of the role of irreducible (closed) sets in domain theory has also been underscored in the solution of the Ho-Zhao problem in~\cite{hojungxi16}.  All these indicate a need to carry out an in-depth and systematic enactment of the scientific program proposed by Jimmie Lawson (as described in the introduction) via our present \emph{replacement} strategy.  A significant part of our research objective is to see how much of domain theory can be developed in the more general setting of topological spaces.

The main result we report herein characterises those $k$-bounded spaces whose $\Irr$-convergence class $\mc{I}$ is topological.  The fundamental property that $k$-bounded spaces $X$ are invariant under the coarsening operator $SI$ plays a key role in the many major arguments employed herein.  The requirement of $k$-bounded sobriety seems indispensable in view that sets of the form $\dua x$ need not be $\tau$-open in an $\Irr$-continuous space $(X,\tau)$.  The present work can be seen as a preliminary investigation of $k$-bounded sober spaces which were first introduced in~\cite{zhaoho15}.  Indeed, $k$-bounded sober spaces deserve a more detailed study in the near future.


\end{document}